\documentclass[11pt, letterpaper]{article}
\usepackage{amsfonts,amsmath,amsopn,amssymb,amsthm,bbm,latexsym,mathrsfs,pstricks,verbatim}

  \usepackage{mathptm}
\let\n\noindent

\font\small=cmr8

\font\tenmsy=msbm10
\font\sevenmsy=msbm10 at 7pt
\font\fivemsy=msbm10 at 5pt
\newfam\msyfam 
\textfont\msyfam=\tenmsy
\scriptfont\msyfam=\sevenmsy
\scriptscriptfont\msyfam=\fivemsy

\let\s\sigma
\let\g\gamma

\let\R\rangle
\let\l\left
\let\r\right

\def\y{{\infty}}

\def\l{{\left}}
\def\r{{\right}}

\def\rw{\rightarrow}

\def\R{\rangle}
\def\frac#1#2{{#1 \over #2}}
\let\La=\Lambda

\def\z{{\cal Z}}

\let\rw\rightarrow
\let\l\left
\let\r\right
\let\s\sigma

\let\la\lambda
\def\M{{\cal M}}

\def\LP{{\rm LP}^{[k-1]}}
\def\rsos{{\rm RSOS}^{[k]}}
\def\q{{\hat q}}
\def\xrw{\xrightarrow}
  \def\exrw{\xrightarrow{{\rm ex} }}

\def\x{{\tilde x}}
\def\w{{\tilde w}}

\font\small=cmr8
\begin{document}

\vskip18pt

\title{\vskip60pt {\bf Paths for  $\z_k$ parafermionic models }}

\vskip18pt


\smallskip
\author{ \bf{P. Jacob and P.
Mathieu}\thanks{patrick.jacob@durham.ac.uk,
pmathieu@phy.ulaval.ca. }
\\
Department of Mathematical Sciences, \\University of Durham, Durham, DH1 3LE, UK\\
and\\
D\'epartement de physique, de g\'enie physique et d'optique,\\
Universit\'e Laval,
Qu\'ebec, Canada, G1K 7P4.
}

\vskip .2in
\bigskip
\date{February 2007}

\maketitle


\vskip0.3cm

\begin{abstract}

We present a simple  bijection  between restricted Bressoud lattice paths  and RSOS paths in regime II of the Andrews-Baxter-Forrester model.  Both types of paths describe states in $\z_k$ parafermionic irreducible modules. The bijection  implies  a direct  correspondence between a RSOS path and a parafermionic state in a  quasi-particle basis.

\end{abstract}
\vskip18pt


\section{Introduction}

\subsection{The two path descriptions of the parafermionic $\z_k$ models}

The analysis of the  Andrews-Baxter-Forrester RSOS model by the corner-transfer-matrix method has led to the  expression of the local-height probability in  terms of a certain one-dimensional  configuration sum \cite{ABF}. Every configuration appearing in the sum is naturally associated to a (RSOS) path. In the infinite-length limit,  this path is in correspondence with  a state of an  irreducible module (specified by the boundary conditions on the path) of  some conformal field theory, and the sum over all paths (with fixed boundary conditions) reproduces  the character of this module \cite{Kyoto, KyotoR}. In regime III, these are the characters of the minimal  models $\M(k+1,k+2)$ \cite{BPZ,CFT}, while  those of the parafermionic $\z_k$ models \cite{ZF} are recovered in regime II. The two regimes are distinguished by the way the paths are weighted. The parameter $k$ specifying these models is related to the original parameter $r$ of \cite{ABF} by $r=k+2$ \cite{Huse}.

The path representation leads naturally to a fermionic form of the characters. The fermionic evaluation of the RSOS configuration sums has been worked out in \cite{Olea,Oleb} in their finitized version. From this,  the fermionic form of the irreducible characters of the minimal models and (by a duality transformation to be reviewed below) those of the parafermionic  models are recovered by taking the  infinite-path limit.  

Somewhat remarkably, there is another path description for the  parafermionic states: these states can be represented by the $(k-1)$-restricted lattice paths described in \cite{BreL}. The most direct way of seeing this relationship is to notice that these paths are actually in one-to-one correspondence with partitions $(\nu_1,\cdots,\nu_m)$ (where $m$ is the total length of the path, to be  defined below) satisfying the following condition \cite{Bur,BreL,JMmu}:
\begin{equation}\label{dis2}
\nu_i-\nu_{i+k-1}\geq 2\;.
\end{equation} 
But this condition  is precisely the combinatorial constraint governing the  quasi-particle basis of states of the  parafermionic  $\z_k$ models \cite{LP,JM}. The conformal dimension of a parafermionic state is the  weight of the corresponding path (also defined below), up to a prescribed correction term  accounting for the fractional dimension of the  parafermionic modes.


The parafermionic states can thus  be described by two quite different  types of paths. Here we show that these paths are essentially the same. In other words, we provide a very simple bijection between the RSOS and Bressoud lattice paths. Since the latter are in one-to-one correspondence with states in a quasi-particle description of the parafermionic modules, we end up with  a similar  correspondence between a RSOS path in regime II and a parafermionic state in a  quasi-particle basis.


\subsection{A duality transformation for RSOS paths}

We recall in this section the duality transformation that relates the $\M(k+1,k+2)$  and the $\z_k$ models within the framework of their RSOS description \cite{ABF,BM}. This duality allows us to translate results derived in the context of the unitary minimal models \cite{Olea,Oleb,FWa}  to the parafermionic case.

The configurations referred to in the terminology  `configuration-sum' are those of  height variables $\s_i\in \{0,1,\cdots ,k\}$, where $i$ ranges from 0 to $L$. Adjacent heights  are subject to the restriction $|\s_i-\s_{i+1}|=1$. The two boundary values are fixed:  let us set for definiteness $\s_0=\s_L=0$. The configuration-sum then takes the form
\begin{equation}\label{cfig}
X(q)= \sum_{\substack{ \s_1,\cdots , \s_{L-1}=0\\|\s_i-\s_{i+1}|=1\\ \s_0=\s_L=0}}^k  q^{\sum_{i=1}^{L-1} \w(i)}\;. 
\end{equation} 
If we plot all the vertices $(i,\s_i)$ of a given configuration and link adjacent vertices by an edge, we obtain  a path,  called  a RSOS path.
The weight function $\w(i)$ depends upon the regime under consideration. We have:
\begin{align}
&{\rm if}~i~{\rm~is~not~an~extremum~of~the~path} : &  \w_{{\rm III}}(i)  = \frac{i}2\quad  {\rm and}\quad  & \w_{{\rm II}}(i)  = 0\;, \nonumber \\
&{\rm if}~i~{\rm~is~an~extremum~of~the~path}:  &  \w_{{\rm III}}(i)  =0  \quad  {\rm and}\quad   & \w_{{\rm II}}(i)  =  \frac{i}2\;.
\end{align} 
The two weight functions are thus related by
\begin{equation}\label{cfig}
 \w_{{\rm III}}(i)  = \frac{i}2- \w_{{\rm II}}(i)\;. 
\end{equation} 
This implies that
\begin{equation}\label{cfig}
X_{{\rm III}}(q)= q^{L(L-1)/4}  \; X_{{\rm II}}(q^{-1})\;. 
\end{equation}
This is the announced duality: up to an irrelevant factor, the two configuration sums are related by $q\rw q^{-1}$ \cite{ABF, BM, Olea, Oleb, FWa}. An operation that increases the weight in one case, decreases it in the dual case. In particular, the minimal-weight configuration corresponding  to the finitized  $\M(k+1,k+2)$  model is mapped to the maximal-weight configuration for the finitized $\z_k$ model.

\subsection{Outline}

The bijection  between lattice paths  and RSOS paths is worked out in various steps in Sec. 2. To keep the presentation as simple as possible, this correspondence is first presented  for paths describing  the vacuum parafermionic module. The bijection between a RSOS path and a parafermionic state entails a new expression for the weight of a RSOS path. An alternative and  direct proof of this novel weight formula is given in Sec.~3.1. The rest of this  section  is concerned with a simple construction of the  generating function of the RSOS path for  a fixed length, that is, a derivation of the finitized parafermionic   vacuum character. The formulae describing a generic module are presented in the final section.

\section{Bijection between lattice paths  and RSOS paths}

\subsection{The multiple-partition  basis}


In \cite{JM.A} (see also \cite{Geo2}), a new quasi-particle-type basis for the  states describing the irreducible modules of the $\z_k$ models has been presented. It is formulated in terms of the modes of the $k-1$ different parafermionic fields $\psi^{(j)}$, with $1\leq j\leq k-1$, of conformal dimension \cite{ZF}:
\begin{equation} h_{\psi^{(j)}}= \frac{j(k-j)}{k}.
\end{equation}   The  modes of $\psi^{(j)}$ are denoted by   $A^{(j)}_{-n+j(j+q)/k}$, where $n$ is an integer. In this notation, the conformal dimension of $A^{(j)}_u$ is $-u$. The fractional part of the mode, $j(j+q)/k$ involves the parafermionic charge $q$ of the state on which the mode acts \cite{ZF}. The charge of $A_u^{(j)}$ is normalized to $2j$ and that of the vacuum state is 0.

Consider for simplicity the states in the vacuum module. These are described by the set of all strings of modes ordered as follows: 
\begin{equation}\label{string} A^{(1)}_{-\la^{(1)}_1+\frac{(1+\q)}{k}}  \cdots A^{(1)}_{-\la^{(1)}_{m_1}+\frac{(1+\q)}{k}} \cdots A^{(k-1)}_{-\la^{(k-1)}_1+\frac{(k-1)(k-1+\q)}{k}} \cdots A^{(k-1)}_{-\la^{(k-1)}_{m_{k-1}}+ \frac{(k-1)(k-1+\q)}{k}} \; | 0\R\;,
\end{equation} 
where $\q$ is an operator which gives the charge of the string at its right. For instance,
\begin{equation}
\q(A_u^{(i)}A_v^{(j)}|0\R) = 2(i+j)+0.
\end{equation} 
It is not difficult to sum the contribution of the fractional part of the modes to the total conformal dimension of  such a string. It actually depends only upon its  total charge  and not its particular composition \cite{JM1}. It can thus be evaluated quite simply by replacing all higher-charge modes $A^{(j)}$ with $j>1$ by $j$ copies of $A^{(1)}$.
The resulting string has then a total  of $m$  operators of type $A^{(1)}$, with
\begin{equation}\label{defm}
m= \sum_{j=1}^{k-1}\, j\, m_j\;,
\end{equation} 
where $m_j$ is the total number of modes $A^{(j)}$ in the original string.
The total `fractional dimension' is then
\begin{equation}\label{fracR}
h_{{\rm frac}}=  \sum_{i=0}^{m-1}\frac{(1+2i)}{k} = \frac{m^2}{k}.
\end{equation}
The dimension of the state (\ref{string}) is 
\begin{equation}
h= \sum_{j=1}^{k-1}\sum_{l=1}^{m_j} \, \la_l^{(j)} - h_{{\rm frac}}.
\end{equation} 

Let us return to (\ref{string}): we still need to specify the constraints on the `integral part' of the modes, the $\la^{(j)}_l$, for these states to form a basis. These conditions are \cite{JM.A}:
\begin{equation} \label{diC}
\la^{(j)}_l\geq \la^{(j)}_{l+1} + 2j\;,  
\end{equation} 
 as well as
\begin{equation}\label{loB}
   \la^{(j)}_{m_j} \geq j+
2j (m_{j+1}+\cdots +  m_{k-1})\;. 
\end{equation} 
 
Since the mode label $\la^{(j)}_l$ keeps track of the charge of the corresponding parafermionic mode via its upper-index and the fractional part of the corresponding mode is uniquely recovered  from the position of the  parafermionic mode in the string (which is itself fixed by the two labels $l$ and $j$), one can rewrite the sequence (\ref{string}) more compactly  as
\begin{equation}\label{strings}  
 \la^{(1)}_1 \cdots  \la^{(1)}_{m_1} \, \la^{(2)}_2\cdots  \la^{(2)}_{m_2} \cdots \la^{(k-1)}_1 \cdots \la^{(k-1)}_{m_{k-1}}\;. 
\end{equation} 
This sequence  is equivalent to the set of $k-1$ ordered partitions of respective lengths $m_1,\cdots , m_{k-1}$, that is, 
\begin{equation}\label{mup} \La^{[k-1]} = (\la^{(1)},\la^{(2)}, \cdots ,\la^{(k-1)})\qquad {\rm with}\qquad 
\la^{(j)}= (\la^{(j)}_1, \cdots , \la^{(j)}_{m_j})\;  .
\end{equation} 
Note that the $m_j$ are allowed to be zero.


\subsection{($k-1)$-restricted lattice paths}

A $(k-1)$-restricted integer lattice path is defined as a sequence of vertices at integral points $(x,y)$ within the strip $x\geq 0$ and $0\leq y\leq k-1$, with adjacent vertices linked by NE, SE or WE (i.e., horizontal) edges, the last one being allowed only if it lies on the $x$-axis \cite{BreL}. There is no definite notion of length for such a path but the final vertex is forced to be  on the $x$-axis. For lattice paths pertaining to the vacuum module, we also require the initial  vertex to be on the $x$-axis.
 
\begin{figure}[ht]
\caption{{\footnotesize An example of lattice path for $k=4$.}} \label{figure1}
\begin{center}
\begin{pspicture}(0,0)(15.5,4)
\psline{->}(0.5,0.5)(0.5,3.5) \psline{->}(0.5,0.5)(15.5,0.5)
\psset{linestyle=dashed,dashadjust=false}
\psline(0.5,2.0)(15.5,2.0)
\psset{linestyle=dotted}
\psline{<->}(4.5,1.0)(6.5,1.0)
\psset{linestyle=solid}
\psline{-}(0.5,0.5)(0.5,0.6) \psline{-}(1.0,0.5)(1.0,0.6)
\psline{-}(1.5,0.5)(1.5,0.6) \psline{-}(2.0,0.5)(2.0,0.6)
\psline{-}(2.5,0.5)(2.5,0.6) \psline{-}(3.0,0.5)(3.0,0.6)
\psline{-}(3.5,0.5)(3.5,0.6) \psline{-}(4.0,0.5)(4.0,0.6)
\psline{-}(4.5,0.5)(4.5,0.6) \psline{-}(5.0,0.5)(5.0,0.6)
\psline{-}(5.5,0.5)(5.5,0.6) \psline{-}(6.0,0.5)(6.0,0.6)
\psline{-}(6.5,0.5)(6.5,0.6) \psline{-}(7.0,0.5)(7.0,0.6)
\psline{-}(7.5,0.5)(7.5,0.6) \psline{-}(8.0,0.5)(8.0,0.6)
\psline{-}(8.5,0.5)(8.5,0.6) \psline{-}(9.0,0.5)(9.0,0.6)
\psline{-}(9.5,0.5)(9.5,0.6) \psline{-}(10.0,0.5)(10.0,0.6)
\psline{-}(10.5,0.5)(10.5,0.6) \psline{-}(11.0,0.5)(11.0,0.6)
\psline{-}(11.5,0.5)(11.5,0.6) \psline{-}(12.0,0.5)(12.0,0.6)
\psline{-}(12.5,0.5)(12.5,0.6) \psline{-}(13.0,0.5)(13.0,0.6)
\psline{-}(13.5,0.5)(13.5,0.6) \psline{-}(14.0,0.5)(14.0,0.6)
\psline{-}(14.0,0.5)(14.0,0.6)
\psline{-}(14.5,0.5)(14.5,0.6)\rput(1.5,0.25){{\small $2$}}
\rput(3.5,0.25){{\small $6$}} \rput(5.5,0.25){{\small $10$}}
\rput(7.5,0.25){{\small $14$}}\rput(10.5,0.25){{\small $20$}}
\rput(13.5,0.25){{\small $26$}}
 \psline{-}(0.5,1.0)(0.6,1.0)
\psline{-}(0.5,1.5)(0.6,1.5) \psline{-}(0.5,2.0)(0.6,2.0)
\psline{-}(0.5,2.5)(0.6,2.5) \psline{-}(0.5,3.0)(0.6,3.0)
\rput(0.25,1.5){{\small $2$}} \rput(0.25,2.5){{\small $4$}}
\psline{-}(0.5,0.5)(1.0,0.5) \psline{-}(1.0,0.5)(1.5,1.0)
\psline{-}(1.5,1.0)(2.0,0.5) \psline{-}(2.0,0.5)(2.5,1.0)
\psline{-}(2.5,1.0)(3.0,1.5) \psline{-}(3.0,1.5)(3.5,2.0)
\psline{-}(3.5,2.0)(4.0,1.5) \psline{-}(4.0,1.5)(4.5,1.0)
\psline{-}(4.5,1.0)(5.0,1.5) \psline{-}(5.0,1.5)(5.5,2.0)
\psline{-}(5.5,2.0)(6.0,1.5) \psline{-}(6.0,1.5)(6.5,1.0)
\psline{-}(6.5,1.0)(7.0,0.5) \psline{-}(7.0,0.5)(7.5,1.0)
\psline{-}(7.5,1.0)(8.0,0.5) \psline{-}(8.0,0.5)(8.5,0.5)
\psline{-}(8.5,0.5)(9.0,0.5) \psline{-}(9.0,0.5)(9.5,0.5)
\psline{-}(9.5,0.5)(10.0,1.0)
 \psline{-}(10.0,1.0)(10.5,1.5)\psline{-}(10.5,1.5)(11.0,1.0)
\psline{-}(11.0,1.0)(11.5,0.5) \psline{-}(11.5,0.5)(12.0,0.5)
\psline{-}(12.0,0.5)(12.5,0.5) \psline{-}(12.5,0.5)(13.0,0.5)
\psline{-}(13.0,0.5)(13.5,1.0) \psline{-}(13.5,1.0)(14.0,0.5)

\end{pspicture}
\end{center}
\end{figure}
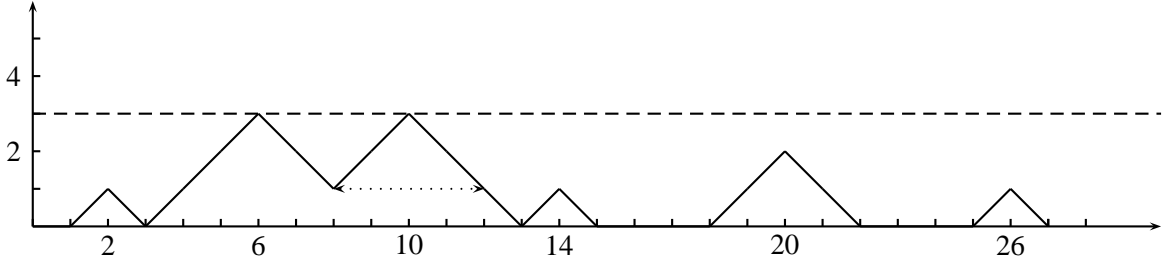

The weight $w$ of a lattice path is  the sum of the $x$-coordinate of all its  peaks:
\begin{equation} \label{defwpath} w= \sum_{x=1}^{L-1} w(x)\qquad {\rm where} \qquad w(x)=
\begin{cases}
x & \text{if $x$ is the position of a peak} ,\\
 0 &\text{ otherwise}\;. \end{cases}
\end{equation}
  The charge (called the relative height in \cite{BreL}) of 
a peak with coordinates $(x,y)$ is the largest integer $c$ such that we can find two vertices $(x',y-c)$ and $(x'',y-c)$ on the path  with $x'<x<x''$ and such that between these two vertices there are no peaks of height larger than $y$ and every peak of height equal to $y$ has weight larger than $x$ \cite{BP}. The charge of a peak is bounded by $1\leq j\leq k-1$. For instance, in Fig. \ref{figure1}, where $k=4$, the charge of the peaks, read from left to right, are 1, 3, 2, 1, 2 and 1. A dotted line indicates the line from which the height must be measured to give the charge. Another  example is provided by Fig. \ref{figure2} (with now $k=5$). There again, dotted lines provide a direct evaluation  of the charges.
The total charge of a path is the sum of the charges of all its peaks.

Such a  path is completely  specified  by the sequence of its peaks, together with their respective charges.
  To prepare the ground for the relation between a path and a parafermionic state, we will understand that the sequence of peaks in a path is to be  read from right to left. The peak data  $(x_n,c_n)$, where $x_n$ is the $x$-coordinate and $c_n$ the charge,  will be  written more  compactly in the form $x_n^{(c_n)}$. We will also refer to $x_n^{(c_n)}$ as a cluster of charge $c_n$ and weight $x_n$. For example, the path of in Fig. \ref{figure1}
corresponds to the sequence $ 26^{(1)} \,20^{(2)}\,14^{(1)}\,10^{(2)}\,6^{(3)}\,2^{(1)} $.

 The basic  characteristics of a path are captured by the following conditions. The first is that a peak of charge $j$ has minimal $x$-coordinate $ j$. The second, referred to as the `path condition', is the following \cite{JMmu}.
  If between two  peaks $ x^{(i)}$ and $ {x'}^{(j)}$ there are peaks all with charge lower than ${\rm min}\, (i,j)$ and whose total charge sums to $c$, then
\begin{equation}\label{dist} x- x' \geq r_{ij}+\chi_{i>j}+2c\;, \end{equation} 
where $ r_{ij} $ stands for
\begin{equation} \label{rij}
 r_{ij} = 2\, {\rm min} \; (i,j)\;, 
 \end{equation}
 and $\chi_{b} =1$ if $b$ is true and 0 otherwise. The case $c=0$ describes the minimal separation between two adjacent peaks: if  $i>j$, it is $2j+1$, while if $i\leq j$, it is $2i.$  Summing up,  a $(k-1)$- restricted lattice path, denoted as $\LP$, is  an ordered sequence of $s$ clusters,
 \begin{equation} \LP =   x_1^{(c_1)}\cdots x_s^{(c_s)} \;, \end{equation}
satisfying:
 \begin{equation} 
 1\leq c_n\leq k-1,\qquad  x_s\geq c_s,\qquad x_n> x_{n+1} ,\qquad \text{and (\ref{dist})}.
 \end{equation}

 In the cluster terminology, a multiple partition like (\ref{mup}) also has the form of an ordered sequence of clusters, but ordered in increasing value of the charge.  This suggests that a multiple partition is nothing but the rearrangement of the sequence of clusters defining a path.  But this calls for a reordering  rule for clusters. For this, we introduce a formal  exchange operation that describes the interchange of two adjacent clusters $ x^{(i)} {x'}^{(j)}$ \cite{JMmu}:
\begin{equation} \label{com}
 x^{(i)} {x'}^{(j)} \quad \exrw \quad ({x'}+r_{ij}) ^{(j)} (x-r_{ij})^{(i)} \;, 
 \end{equation}
 where $r_{ij}$ is defined in (\ref{rij}).
This  operation preserves the individual values of the charge and also the sum of the  weights. 
Note that after a number of interchanges on a sequence of clusters  $\{ x_n^{(c_n)} \}$ describing a path, such that $\{ x_n^{(c_n)} \}\exrw
 \{ {\x_n}^{(c_n)} \}$, the values  $\x_n$ are no longer necessarily decreasing and they no longer  correspond to peak positions in a modified path.  Indeed, (\ref{com}) typically spoils the path condition (\ref{dist}). 
 


The correspondence  between a $(k-1)$-restricted lattice path $\LP$ and a multiple partition $ \La^{[k-1]}$ is now  easily described \cite{JMmu}. Take a path read from right to left and use the exchange relation (\ref{com}) to reorder the clusters in increasing value of the  charge from left to right.  The weights of the resulting clusters of  charge $j$ then form the parts of the partition $\la^{(j)}$. The multiple partition is thus a  canonical rewriting of the original path, with 
\begin{equation} 
s= \sum_{j=1}^{k-1} m_j\;.
\end{equation} 
  This gives the link 
$  \LP \rw \La^{[k-1]}.$
 For instance, we have
\begin{equation}  \label{exB}
\, 26^{(1)} \,20^{(2)}\,14^{(1)}\,10^{(2)}\,6^{(3)}\,2^{(1)} \exrw
 26^{(1)}\, 16^{(1)}\, 8^{(1)}\, 16^{(2)}\,8^{(2)}\, 4^{(3)}  \;. \end{equation} 
The multiple partition thus obtained is
\begin{equation}  \label{exBa}
  \la^{(1)} = (26,16,8),\quad  \la^{(2)} = (16,8),\quad \la^{(3)} = (4).
\end{equation} 

The inverse operation,  $ \La^{[k-1]} \rw  \LP $, amounts to viewing   $(\la^{(1)},\cdots,\la^{(k-1)})$ as a sequence of clusters (cf. (\ref{strings})) and reordering the clusters in decreasing value of the weight using (\ref{com})  to ensure that the condition (\ref{dist}) is  everywhere satisfied. As shown in \cite{JMmu}, this is a finite process and it  has  a unique solution. We have thus a bijection
\begin{equation}\label{bi1}
 \LP \leftrightarrow    \La^{[k-1]}.
\end{equation}

Let us return briefly to the parafermionic interpretation of the lattice path. A state of the form (\ref{string}) corresponds to a multiple partition $\La^{[k-1]}$ and thus to a lattice  path $\LP$. A cluster of charge $j$ is  thus a combinatorial representation of a parafermionic mode  $A^{(j)}$ and the parafermionic charge is twice the cluster charge. Moreover, if the path has total charge $m$ and weight  $w$, the conformal dimension of the corresponding parafermionic state is
\begin{equation} h= w-h_{{\rm frac}}= w-\frac{m^2}{k}\;.
\end{equation}
Note that the exchange 
relation (\ref{com}) is an abstract version of the parafermionic mode commutation relations. Recall that these commutations take the form of infinite sums (the so-called generalized commutations relations) \cite{ZF}. The relation (\ref{com})  amounts,  roughly, to considering only the leading term of the two infinite strings of states. A similar relation is introduced in \cite{BreL} and called a shuffle.

\subsection{From the multiple partition  to the  RSOS path}

Let us now describe a simple operation that transforms a set of $(k-1)$ ordered  partitions into a set of $k$ ordered ones by adding clusters of charge $k$ in a suitable way. The procedure is to 
add a sequence of   $m_k$ closely-packed clusters of charge $k$ (with prescribed parts) to the left  of the sequence described by $\La^{[k-1]}$ and  move them to the right of $\La^{[k-1]}$, in order for the clusters to be ordered in increasing value of the charge. The number  $m_k$ is   determined below. Let us denote the weight of the added clusters, once in their rightmost position, by $\gamma_l^{(k)}$. The closely-packed condition refers to the condition (\ref{diC}) with the equality sign: $\g^{(k)}_l=\g^{(k)}_{l+1}+2k$.  We have thus moved through $\La^{[k-1]}$ a partition with fixed parts, which, once in rightmost position, are given by
\begin{equation} \label{defg}
\g^{(k)} = ((2m_k-1)k,\cdots , 3k,k) \;. \end{equation}
(It should be clear that this fixes the value of the parts inserted at the initial step, before they get  displaced to the right of $\La^{[k-1]}$). Because the added charge $k$ clusters have been moved at the right of $\La^{[k-1]}$, we need to reshuffle all the parts  of $\La^{[k-1]}$ according to the exchange relation (\ref{com}), so that:
\begin{equation} \label{boost}
\la^{(j)} _l\rw \g^{(j)} _l = \la^{(j)} _l+2jm_k\;.
\end{equation}
The conditions on these reshuffled  parts are thus 
\begin{equation} \label{diCg}
\g^{(j)}_l \geq \g^{(j)}_{l+1} + 2j\;, 
\end{equation} and 
\begin{equation}\label{loBg}
   \g^{(j)}_{m_j} \geq j+
2j (m_{j+1}+\cdots +  m_{k}).
\end{equation} 
For $j=k$, these relations  are satisfied as equalities.
Denote the resulting set of $k$ ordered partitions as $\Gamma^{[k]}$:
\begin{equation}\label{Gk}
 \Gamma^{[k]}= (\g^{(1)},\cdots ,  \g^{(k)}), \qquad {\rm with}\qquad 
\g^{(j)}= (\g^{(j)}_1, \cdots , \g^{(j)}_{m_j}).
\end{equation}
However, for this definition to be complete, we still need to fix $m_k$. 
Once this value  will be determined, we will have a unique procedure to go from $\La^{[k-1]}$ to $\Gamma^{[k]}$ and back.  That will establish the bijection
\begin{equation}\label{bi2}
 \La^{[k-1]}  \leftrightarrow  \Gamma^{[k]}.
\end{equation}


Let us now transform $ \Gamma^{[k]}$ into a $k$-restricted RSOS path, denoted  $\rsos$. The procedure is exactly the very one  used to transform $\La^{[k-1]} $ into a lattice path $\LP$. This is the point where the number $m_k$ gets fixed. It is determined by enforcing that the resulting path is connected, that is, to be such that all peaks are in contact in the sense that nowhere adjacent peaks get separated by a horizontal edge.
We thus fix $m_k$ by requiring that the modified path (which now has peaks of charge $k$) is completely  free of horizontal segments. 
This can always be ensured by the insertion of sufficiently many clusters of charge $k$ since the size of each added cluster, $2k$, is larger than the increase in the weight of a cluster of charge $j$, which is $2j$, that results from  the insertion
of each charge $k$-cluster.

Let us now see how this can be made precise.
Consider the substring containing only the clusters of charge $ j$ up to $k$. We then reorder the  clusters in decreasing value of the weights by also ensuring that the path condition is everywhere satisfied. In addition, we require the rightmost peak of charge $j$ to lie on a connected path. 
The length of a RSOS path is twice its charge content (and the charge of the peaks in a RSOS path can be read off exactly as for a lattice path \cite{Olea}). Since the charge content of the path containing only peaks of charge $\geq j$ is 
\begin{equation} 
jm_j+(j+1)m_{j+1}+\cdots +km_k\; , 
\end{equation} 
the connecting criterion for the rightmost peak of charge $j$  translates into
\begin{equation} \label{upup}
\g^{(j)}_1+j \leq 2\,[\, jm_j+(j+1)m_{j+1}+\cdots +km_k\,]\;.
\end{equation} 
The peak of charge $j$ as a whole is represented by a triangle with peak $x$-position at $\g^{(j)}_1$; this position as well as the following $j$ SE edges of the triangle,  must both lie within the length that is determined by the charge content. This is the meaning  of (\ref{upup}).
If we reexpress this inequality in  terms of the $\La^{[k-1]} $ data, we have
\begin{equation} 
\la^{(j)}_1+j  + 2jm_k \leq 2\,[\, jm_j+\cdots +km_k \,]\;.
\end{equation} 
For any given value of $\la^{(j)}_1$, there is always a minimal value of $m_k$ that ensures this bound. Indeed, the above condition implies
\begin{equation} 
m_k\geq  \l\lceil \frac{\la^{(j)}_1+j-2\,[\, jm_j+\cdots +(k-1)m_{k-1} \,]}{2(k-j)} \r\rceil\;, 
\end{equation} 
where $\lceil x\rceil$ is the smallest integer $\leq x$.
This condition must hold for all values of $j$, from $k-1$ down to 1. We thus set
\begin{equation} \label{delfmk}
m_k= {\rm max}\; \l(  \l\lceil \frac{\la^{(j)}_1+j-2\,[\, jm_j+\cdots +(k-1)m_{k-1} \,]}{2(k-j)} \r\rceil  , 1  \leq j\leq k-1\r )\; .
\end{equation}
This is the announced determination of $m_k$.  This specification completes the proof of (\ref{bi2}).
For the example (\ref{exB})-(\ref{exBa}), we have
\begin{equation} \label{exbc}
m_4= {\rm max}\; \l(  \l\lceil \frac76 \r\rceil  , \l\lceil \frac44 \r\rceil, \l\lceil \frac12 \r\rceil \r )= {\rm max}\; \l( 2,1,1\r) = 2\; .
\end{equation}

This minimal number of inserted peaks of charge $k$ gives rise to the RSOS path of lowest possible length for a given Bressoud lattice path. If the  number of inserted peaks is larger than required by the connectivity criterion, in order to match a prescribed length for instance, the extra peaks of charge $k$ stay at the right of the path.

Inverting the procedure, we can define the multiple partition $\Gamma^{[k]}$ as in (\ref{Gk}), in terms of the conditions (\ref{diCg}), (\ref{loBg}) as well as 
\begin{equation} \label{upg}
\g^{(j)}_1 \leq -j  + 2\,[\, jm_j+(j+1)m_{j+1}+\cdots +km_k\,]\;.
\end{equation} 
These conditions hold for all $j$, including $j=k$. In particular, the lower bound on $\g^{(k)}_{m_k}$, the upper bound on $\g^{(k)}_{1}$, together with the difference condition (\ref{diCg}),  imply  that $\g^{(k)} $ is fixed to (\ref{defg}). This establishes the bijection
\begin{equation} \label{bi3}
\Gamma^{[k]}  \leftrightarrow  \rsos.
\end{equation}
The chain of bijections (\ref{bi1}), (\ref{bi2}) and  (\ref{bi3})  implies the following correspondence:
\begin{equation} \label{bi4}
\LP  \leftrightarrow  \rsos.
\end{equation}
This is our main result. 

In the light of \cite{JMmu}, where $\LP$ is related to a partition satisfying the condition (\ref{dis2}), the correspondence (\ref{bi4}) entails a direct bijection between such a partition and a RSOS path.

We stress that a set of $k$ ordered partitions has an immediate interpretation in terms of parafermionic modes. For instance $\Gamma^{[k]}$ is associated to the following state
\begin{equation}\label{stringext} A^{(1)}_{-\g^{(1)}_1+\frac{(1+\q)}{k}}  \cdots A^{(1)}_{-\g^{(1)}_{m_1}+\frac{(1+\q)}{k}} \cdots A^{(k)}_{-\g^{(k)}_1+\frac{(k)(k+\q)}{k}} \cdots A^{(k)}_{-\g^{(k)}_{m_{k}}+ \frac{(k)(k+\q)}{k}} \; | 0\R\;,
\end{equation} 
The bijection (\ref{bi3})  thus  underlies a direct combinatorial interpretation  of a parafermionic quasi-particle state  in terms of a RSOS path.

\subsection{ Examples}

 \begin{figure}[ht]
\caption{{\footnotesize A RSOS path for $k=4$.}} \label{figure2}
\begin{center}
\begin{pspicture}(0,0)(12.5,3)
\psline{->}(0.3,0.3)(0.3,2.5) \psline{->}(0.3,0.3)(12.0,0.3)
\psset{linestyle=dashed,dashadjust=false}
\psline(0.3,1.5)(12.0,1.5)
\psset{linestyle=dotted}
\psline{<->}(0.6,0.6)(1.2,0.6)
\psline{<->}(4.2,0.6)(4.8,0.6)
\psline{<->}(4.8,0.6)(6.6,0.6)
\psline{<->}(2.7,0.9)(3.9,0.9)
\psline{<->}(8.4,0.6)(9.0,0.6)
\psset{linestyle=solid}
\psline{-}(0.3,0.3)(0.3,0.4) \psline{-}(0.6,0.3)(0.6,0.4)
\psline{-}(0.9,0.3)(0.9,0.4) \psline{-}(1.2,0.3)(1.2,0.4)
\psline{-}(1.5,0.3)(1.5,0.4) \psline{-}(1.8,0.3)(1.8,0.4)
\psline{-}(2.1,0.3)(2.1,0.4) \psline{-}(2.4,0.3)(2.4,0.4)
\psline{-}(2.7,0.3)(2.7,0.4) \psline{-}(3.0,0.3)(3.0,0.4)
\psline{-}(3.3,0.3)(3.3,0.4) \psline{-}(3.6,0.3)(3.6,0.4)
\psline{-}(3.9,0.3)(3.9,0.4) \psline{-}(4.2,0.3)(4.2,0.4)
\psline{-}(4.5,0.3)(4.5,0.4) \psline{-}(4.8,0.3)(4.8,0.4)
\psline{-}(5.1,0.3)(5.1,0.4) \psline{-}(5.4,0.3)(5.4,0.4)
\psline{-}(5.7,0.3)(5.7,0.4) \psline{-}(6.0,0.3)(6.0,0.4)
\psline{-}(6.3,0.3)(6.3,0.4) \psline{-}(6.6,0.3)(6.6,0.4)
\psline{-}(6.9,0.3)(6.9,0.4) \psline{-}(7.2,0.3)(7.2,0.4)
\psline{-}(7.5,0.3)(7.5,0.4) \psline{-}(7.8,0.3)(7.8,0.4)
\psline{-}(8.1,0.3)(8.1,0.4) \psline{-}(8.4,0.3)(8.4,0.4)
\psline{-}(8.7,0.3)(8.7,0.4)
\psline{-}(9.0,0.3)(9.0,0.4)
\psline{-}(9.3,0.3)(9.3,0.4) \psline{-}(9.6,0.3)(9.6,0.4)
\psline{-}(9.9,0.3)(9.9,0.4) \psline{-}(10.2,0.3)(10.2,0.4)
\psline{-}(10.5,0.3)(10.5,0.4) \psline{-}(10.8,0.3)(10.8,0.4)

\rput(0.9,-0.05){{\small $2$}}
\rput(2.1,-0.05){{\small $6$}} \rput(3.3,-0.05){{\small $10$}}
\rput(4.5,-0.05){{\small $14$}}\rput(5.7,-0.05){{\small $18$}}
\rput(7.5,-0.05){{\small $24$}}
\rput(8.7,-0.05){{\small $28$}}\rput(9.9,-0.05){{\small $32$}}
 \psline{-}(0.3,0.6)(0.4,0.6)
\psline{-}(0.3,0.9)(0.4,0.9) \psline{-}(0.3,1.2)(0.4,1.2)
\psline{-}(0.3,1.5)(0.4,1.5) \psline{-}(0.3,1.8)(0.4,1.8)

\rput(0.05,0.9){{\small $2$}} \rput(0.05,1.5){{\small $4$}}
\psline{-}(0.3,0.3)(0.6,0.6) \psline{-}(0.6,0.6)(0.9,0.9)
\psline{-}(0.9,0.9)(1.2,0.6) \psline{-}(1.2,0.6)(1.5,0.9)
\psline{-}(1.5,0.9)(1.8,1.2) \psline{-}(1.8,1.2)(2.1,1.5)
\psline{-}(2.1,1.5)(2.4,1.2) \psline{-}(2.4,1.2)(2.7,0.9)
\psline{-}(2.7,0.9)(3.0,1.2) \psline{-}(3.0,1.2)(3.3,1.5)
\psline{-}(3.3,1.5)(3.6,1.2) \psline{-}(3.6,1.2)(3.9,0.9)
\psline{-}(3.9,0.9)(4.2,0.6) \psline{-}(4.2,0.6)(4.5,0.9)
\psline{-}(4.5,0.9)(4.8,0.6) \psline{-}(4.8,0.6)(5.1,0.9)
\psline{-}(5.1,0.9)(5.4,1.2) \psline{-}(5.4,1.2)(5.7,1.5)
\psline{-}(5.7,1.5)(6.0,1.2)
 \psline{-}(6.0,1.2)(6.3,0.9)\psline{-}(6.3,0.9)(6.6,0.6)
\psline{-}(6.6,0.6)(6.9,0.3) \psline{-}(6.9,0.3)(7.2,0.6)
\psline{-}(7.2,0.6)(7.5,0.9) \psline{-}(7.5,0.9)(7.8,0.6)
\psline{-}(7.8,0.6)(8.1,0.3) \psline{-}(8.1,0.3)(8.4,0.6)
\psline{-}(8.4,0.6)(8.7,0.9) \psline{-}(8.7,0.9)(9.0,0.6)
\psline{-}(9.0,0.6)(9.3,0.9) \psline{-}(9.3,0.9)(9.6,1.2)
\psline{-}(9.6,1.2)(9.9,1.5) \psline{-}(9.9,1.5)(10.2,1.2)
\psline{-}(10.2,1.2)(10.5,0.9) \psline{-}(10.5,0.9)(10.8,0.6)
\psline{-}(10.8,0.6)(11.1,0.3)

\end{pspicture}
\end{center}
\end{figure}
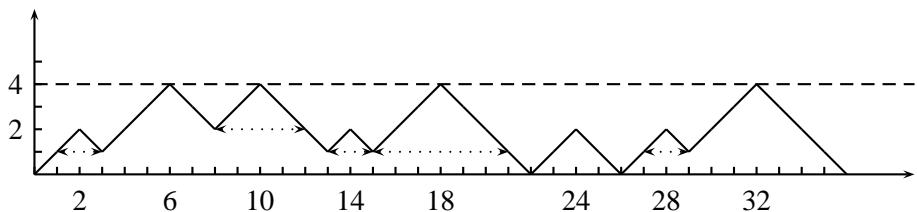

Let us illustrate the map $\rsos\rw\LP$ by considering the $k=4$ RSOS path of Fig. \ref{figure2}.
Its reordering in a multiple partition denoted $\Gamma^{[4]}$ is:
\begin{equation}  
32^{(4)} \,28^{(1)} \,24^{(2)}\,18^{(3)}\,14^{(1)}\,10^{(2)}\,6^{(4)}\,2^{(1)} \,\exrw \,30^{(1)} \,20^{(1)}\,12^{(1)}\,24^{(2)}\,16^{(2)}\,16^{(3)}\,12^{(4)}\,4^{(4)}\;. 
\end{equation} 
Now delete the clusters of charge 4 and reduce the indices by inverting (\ref{boost}).  This gives $\La^{[3]}$.  Then  reorder the clusters to generate a lattice path:
\begin{equation}  
26^{(1)}\, 16^{(1)}\, 8^{(1)}\, 16^{(2)}\,8^{(2)}\, 4^{(3)}\, \exrw \, 26^{(1)} \,20^{(2)}\,14^{(1)}\,10^{(2)}\,6^{(3)}\,2^{(1)}\;, 
\end{equation} 
which is the path of Fig. {\ref{figure1}. This last line is the inverse of (\ref{exB}). Reversing this example, the first step amounts to computing the value of $m_4$: this is done in (\ref{exbc}). 

As a second example, let us relate the lattice  paths of weight 6 for $k=3$ to their corresponding RSOS paths obtained by adding clusters of charge 3 and boosting the weights of the lower charged clusters accordingly (cf. (\ref{boost})). The value of $m_3$ required in each case is calculated from (\ref{delfmk}) and it is indicated in every case. With $m_3$ fixed, the weights of $\La^{[2]}$ are reshuffled appropriately. Then, if necessary, the clusters are reordered with the exchange relation (\ref{com})  in order to respect the path condition (\ref{dist}). The rightmost sequence is the corresponding RSOS path:
\begin{alignat}{3}
& 6^{(1)}\phantom{1^{(1)}}  \quad  \xrw{m_3=2} \quad 10^{(1)} \,9^{(3)}\,3^{(3)} &\quad &\exrw \quad11^{(3)} 8^{(1)}\,3^{(3)}\nonumber \\
& 6^{(2)}  \phantom{1^{(1)}}    \quad  \xrw{m_3=2} \quad14^{(2)}\, 9^{(3)}\,3^{(3)} &&\nonumber \\
& 5^{(1)} \,1^{(1)}   \quad  \xrw{m_3=1}\quad7^{(1)} 3^{(1)}  3^{(3)}  \quad  &&\exrw\quad 7^{(1)} \,5^{(3)}\,1^{(1)} \nonumber \\
& 5^{(2)} \,1^{(1)}  \quad  \xrw{m_3=1}\quad 9^{(2)} \,3^{(1)} \, 3^{(3)}  \quad  &&\exrw\quad9^{(2)} \,5^{(3)}\, 1^{(1)} \nonumber \\
& 4^{(1)}\, 2^{(1)}   \quad  \xrw{m_3=1}\quad 6^{(1)} \,4^{(1)} \, 3^{(3)}  \quad  &&\exrw\quad7^{(3)}\, 4^{(1)}\,2^{(1)} \nonumber \\
& 4^{(1)}\, 2^{(2)}   \quad \xrw{m_3=0}\quad 4^{(1)}\,2^{(2)} \;.  \end{alignat}
In the last case, the lattice path has no horizontal move, hence it remains unchanged.

\section{Weighting and counting RSOS paths}

\subsection{The weight of a RSOS path}

The weight of a RSOS path (with the understanding that our discussion pertains solely to regime II) is  given
in the form
\begin{equation} \label{wdefrsos}
\w= \sum_{x=1}^{L-1}  \w(x)\qquad {\rm where} \qquad \w(x)=\begin{cases}
\frac{x}2 & \text{if $x$ is an extremum of the path} \\
0& \text{otherwise}\;. 
\end{cases} \end{equation}
To make contact with the conformal dimension of a state, we need to subtract from this the contribution of the ground state. The latter, in turn, is specified by the total charge, say $m'$, modulo $k$. If we set:
\begin{equation} 
 m'= pk+r\;, 
\end{equation} 
the ground state is thus characterized by the integer $r$ with $0\leq r\leq k-1$. In the terminology of \cite{JM1}, $2r$ gives  the relative charge of the parafermionic (fixed-charge modulo $2k$) submodule. In other words, the vacuum module can be broken up into $r$ distinct submodules of different relative charge. The dimension of the highest-weight state of (parafermionic) charge $2r$ in the vacuum module is \cite{JM1}:
\begin{equation} 
 h_0^{(r)} = h_{\psi^{(r)}}= \frac{r(k-r)}{k}\;. 
\end{equation} 
The dimensions of the states within a given charged module all differ by integers.

Let us call gs($r$) the path representing the highest-weight state of charge $2r$ in the vacuum module and by path($r$) a path of total (path) charge $m'=pk+r$ for some $p$. The conformal dimension of the corresponding state is   thus
\begin{equation} \label{wrsos}
 h= \w_{{\rm path}(r)}- \w_{{\rm gs}(r)} + h_0^{(r)} \;. 
\end{equation} 

The correspondence between a RSOS path and a parafermionic  state like (\ref{stringext}) 
 a different expression for the conformal dimension. This alternative form, read off from (\ref{stringext}),  is expressed in terms of the sum of the peak $x$-positions, $w$, and the fractional dimension associated to the total charge:
\begin{equation} \label{wbres} h= w_{{\rm path}(r)}-h_{{\rm frac}}= w_{{\rm path}(r)}-\frac{{m'}^2}{k}\; . 
\end{equation} 
We stress that in this expression, we sum over the position of the peaks (cf. (\ref{defwpath}))  and not half these positions and that the minima no longer contribute. Moreover, the weight of the path, corrected by $h_{{\rm frac}}$, yields directly the conformal dimension of the corresponding state. In other words, we no longer need to subtract the ground-state contribution. 

The equivalence of (\ref{wrsos}) and (\ref{wbres}) is a direct consequence of the bijection (\ref{bi4}). However, it is of interest to prove it  directly.

We first consider the path gs($r$), the ground state specified by $r$ (so that the total charge is $m'=pk+r$). It is described by  a peak of charge $r$ at position $r$ followed by $p$ peaks of charge $k$, at positions $k+2r,3k+2r,\ldots , (2p-1)k+2r$.  Its weight  is simply evaluated:
\begin{equation}  w_{{\rm gs(r)}}= r+kp^2+ 2rp\;.
\end{equation} 
The first term corresponds to the weight of the first peak. The second term is $w_{{\rm gs(0)}}$. Finally, the third term describes the modification to the weight $w_{{\rm gs(0)}}$ caused by the shift in the  position of the $p$ peaks, resulting from  the insertion at the beginning of the path, of a peak of height $r$. The expression (\ref{wbres}) yields
\begin{equation} \label{wgs}h= w_{{\rm gs(r)}}-h_{{\rm frac}}= r+kp^2+2rp-\frac{(kp+r)^2}{k}= \frac{r(k-r)}{k}=h_0^{(r)} .
\end{equation} 
This agrees with  the  expression (\ref{wrsos}) in  the case where path$(r)$=gs$(r)$.

In the second step, we observe that any configuration can be obtained from gs($r$) by combining the two elementary operations:

\n 1- {Keep the charge content fixed and displace  a peak of charge $j\leq k-1$ toward the right by one unit.}

\n 2- {Change the charge content by breaking a peak of charge $n$ into two peaks of charge $j$ and $n-j$.}

\n We will verify that the modification in the weight, resulting from these two operations, is the same whether it is computed from  (\ref{wrsos}) or (\ref{wbres}), that is:
\begin{equation} \label{wdif} w-w_{{\rm gs(r)}}=  \w -\w_{{\rm gs(r)}}\;.
\end{equation} 
Obviously, (\ref{wgs}) and (\ref{wdif}) imply directly the equivalence  of  (\ref{wrsos}) and (\ref{wbres}):
\begin{equation} 
h= w_{{\rm path}(r)}-h_{{\rm frac}}=  (w_{{\rm path}(r)}-w_{{\rm gs}(r)}) +w_{{\rm gs}(r)}- h_{{\rm frac}}\nonumber= (\w_{{\rm path}(r)}-\w_{{\rm gs}(r)})+h_0^{(r)}\;. 
\end{equation} 

It only remains to verify the equality (\ref{wdif}) for the two basic operations. For a given charge content, the different configurations that can be generated are all obtained from a minimal-weight configuration (described in Sec. 3.3) through a sequence of basic displacements. These are described in detail in \cite{BreL,Olea}. (In the latter reference, the discussion pertains to the unitary minimal models; it can be applied directly to our case by reversing the direction of the displacements as discussed in Sec. 1.2). Consider then the displacement of a peak by one unit toward the right (operation (1)). The weight $w$ is increased by 1 since the peak position is increased by 1. Similarly, $\w$ changes by 1 since both the peak and the minimum just at its right both increase by 1 (and the sum is divided by 2). Therefore, a combination of such peak displacements always satisfy (\ref{wdif}).

Finally, let us compare the weight of a peak with respect to that of two peaks with same total charge and in-between the same initial and final positions (operation (2)). Consider a peak of height $n$ centered at $n+x$. We have:
\begin{equation} 
(n+x)^{(n)} : \qquad  w= n+x\,,\qquad  \w= \frac12(n+x)\;.
\end{equation} 
 Compare this with the weight  of the path described by a peak of height $j$ at $j+x$ followed by a peak of weight $n-j$  at $n+j+x$: 
\begin{equation} (n+j+x)^{(n-j)}(j+x)^{(j)}  : \qquad  w=  n+2j+2x\qquad  \w= \frac12(n+4j+3x)   \;,
\end{equation} 
so that  
\begin{equation}
\Delta w= \Delta \w=  2j+x \;.
\end{equation} 
 Iterating this procedure, one can break a peak into more than two peaks  of  lower charge and  at each step we will preserve (\ref{wdif}). This completes the proof of the equivalence  of  (\ref{wrsos}) and (\ref{wbres}).

\subsection{Enumerations of the RSOS paths}

An intermediate step toward the derivation of a finitized form of the parafermionic character  amounts to 
counting the number of allowed configurations with fixed values of the $m_j$.  We will relate this enumeration problem to that of counting partitions with simple restrictions. For this, let us subtract from $\g^{(j)}$ the following partition with $m_j$ equal parts:
\begin{equation} 
(\Delta_j, \cdots , \Delta_j ) \qquad {\rm where }\qquad  \Delta_j = j+ 2j\,[\,m_{j+1}+\cdots +m_k\,]\;,
\end{equation} 
as well as the staircase:
\begin{equation} 
((m_j-1)2j,\cdots, 4j,  2j,0)\;. 
\end{equation} 
In other words, we define a new partition $\s^{(j)}$ by 
\begin{equation} 
\s_l^{(j)} =\g_l^{(j)}  -\Delta_j -2j (m_j-l)\;.
\end{equation} 
This implies that $\s^{(j)}_{m_j}\geq 0$. The upper bound (\ref{upg}) yields
\begin{align} 
\s_1^{(j)} &\leq -j  + 2\,[\, jm_j+(j+1)m_{j+1}+\cdots +km_k\,]  -\Delta_j -2j (m_j-1) \nonumber \\ &= 
2\,[\, (j+1)m_{j+1}+\cdots +km_k\,]  - 2j\,[\,m_{j+1}+\cdots +m_k\,] \nonumber \\ &= 
2m_{j+1}+4m_{j+2}+\cdots +2(k-j)m_k \;.
\end{align}
We have thus found that $\s^{(j)}$ is a partition with at most $m_j$ parts and whose parts are all $\leq p_j$ with $p_j$ given by
\begin{equation} 
p_j= 2m_{j+1}+4m_{j+2}+\cdots +2(k-j)m_k.
\end{equation} 
Their enumeration is simply (omitting the upperscript $(j)$ in the summation bounds) \cite{Andr}:
\begin{equation}\label{combi}
\sum_{\s_1=0}^{p_j} \sum_{\s_2=0}^{\s_1}\cdots \sum_{\s_{m_j}=0}^{\s_{m_j-1}}  1 = 
 \begin{pmatrix}
 p_j+m_j\\ m_j\end{pmatrix}\;.
\end{equation} 
The total number of elements $\Gamma^{[k]}$, with all $m_j$ fixed, is thus
\begin{equation}\label{combiT}
\sum_{\s^{(1)}}\sum_{\s^{(2)}} \cdots \sum_{\s^{(k-1)}}  1 = 
 \prod _{j=1}^{k-1} \begin{pmatrix}
 p_j+m_j\\ m_j\end{pmatrix}\;.
\end{equation}
Recall that  the partition $\s^{(k)}$ has zero parts and it is thus unique (so that it is not summed over).

For instance, consider 
$k=3$ and  the charge content $(m_1,m_2,m_3)=(1,2,1)$, so that $\Gamma^{[3]}=
\g^{(1)}_1 \, \g^{(2)}_1\, \g^{(2)}_2\,  \g^{(3)}_1. $
The bounds (\ref{diCg}), (\ref{loBg}) and (\ref{upg}) give:
\begin{equation} 
7 \leq \g^{(1)}_1\leq 15\;, \quad \g^{(2)}_2\geq 6\;,  \quad \g^{(2)}_1\leq 12\;,\quad \g^{(2)}_1\geq\g^{(2)}_2+4\;, \quad {\rm  and} \quad 3\leq \g^{(3)}_1\leq 3.
\end{equation}
There are 9 choices for $\g^{(1)}_1$, 6 possible values of $(\g^{(2)}_1, \g^{(2)}_2)$ and $\g^{(3)}_1$ is fixed. That makes a total of 54 possibilities, in agreement with the combinatorial factor (\ref{combiT}):
\begin{equation} 
\begin{pmatrix}
m_1+2m_2+4m_3\\ m_1\end{pmatrix}\begin{pmatrix}
m_2+2m_3\\ m_2\end{pmatrix}= \begin{pmatrix} 9\\ 1\end{pmatrix}\begin{pmatrix}
4\\ 2\end{pmatrix}= 54.
\end{equation}



\subsection{The finitized fermionic character}

The generating function of all RSOS paths, which corresponds to a finitization of the $\z_k$  parafermionic vacuum character, is obtained by first computing the weight of the minimal-weight configuration for a fixed charge content, evaluating  the weight of the  different configurations that can be obtained from this one (the number of such configurations is given by (\ref{combiT}))  by the allowed displacements of the peaks,  and finally by summing over all values of $m_j$ compatible with a fixed value of the total charge (that is, a fixed length).  The analysis is quite similar to that of \cite{Olea} for the unitary minimal models and in consequence, the presentation here will be somewhat succinct.

The vacuum minimal-weight configuration (mwc(0)) for a fixed charge content (fixed values of the $m_j$)
is the one for which all peaks are isolated (this means that their base vertices lie on the horizontal axis but, of course, they must be closely packed as suited for a RSOS path) and ordered by  increasing value of the charge from left to right. The weight is easily found to be
\begin{equation}  w_{{\rm mwc(0)}} = m_1^2+2m_1m_2+2m_2^2+\cdots + [2m_1+\cdots +2(k-1)m+{k-1}] m_k+km_k^2\;, 
\end{equation} 
or equivalently
\begin{equation}  w_{{\rm mwc(0)}} =\frac12 \sum_{i,j=1}^k r_{ij} \, m_i m_j\;, 
\end{equation} 
with $r_{ij}$ defined in (\ref{rij}).
The conformal dimension of this configuration is thus 
\begin{equation}
h_{{\rm mwc(0)}} =  w_{{\rm mwc(0)}} - \frac{{m'}^2}{k} = 
   \frac12 \sum_{i,j=1}^k r_{ij} \, m_i m_j -  \frac{(m+km_k)^2}{k}=
 \frac12 \sum_{i,j=1}^{k-1} r_{ij} \, m_i m_j -  \frac{m^2}{k}.
\end{equation}   
Note the cancellation of $m_k$ in the last step: the conformal dimension of  the minimal-weight configuration  is thus  independent of the number of  modes of type $k$.

Any configuration with the same charge content can be obtained from  the minimal-weight configuration by a sequence of unit displacements of peaks of charge $<k$ toward the right. Each unit move increases the weight by 1.  Because the length of the path is fixed, the number of such displacements is  bounded: the maximal displacement of a peak of charge $j$ is $p_j$. Taken into account,  the weight modification transforms  the   binomial coefficient (\ref{combi}) into a  $q$-binomial \cite{Andr}:
\begin{equation}\label{combi}
\sum_{\s_1=0}^{p_j} \sum_{\s_2=0}^{\s_1}\cdots \sum_{\s_{m_j}=0}^{\s_{m_j-1}}  q^{\s_1+\cdots +\s_{m_j} } = 
 \begin{bmatrix}
 p_j+m_j\\ m_j\end{bmatrix}\;.
\end{equation} 
where
 \begin{equation} \begin{bmatrix}
m\\ n\end{bmatrix}= \frac{(q)_m}{(q)_n(q)_{m-n}}\qquad {\rm with}\qquad (q)_n= \prod_{i=1}^n (1-q^i)\;.
\end{equation} 
Therefore, the product of   binomial coefficients (\ref{combiT}) is transformed into the product of $q$-binomials:
\begin{equation} \prod _{j=1}^{k-1} \begin{pmatrix}
 p_j+m_j\\ m_j\end{pmatrix}\quad \rw \quad \prod _{j=1}^{k-1} \begin{bmatrix}
 p_j+m_j\\ m_j\end{bmatrix}\;,
 \end{equation}

The finitized (i.e., for $m'$ finite) vacuum character is obtained by multiplying  the above number of $q$-weighted (relative to mwc(0)) configurations, incorporate the weight of the minimal-weight configuration and sum over all values of $m_j$ compatible with $\sum_{j=1}^k jm_j = m'$:
 \begin{equation}
 \chi^{(m')}_0(q) = \sum_{\substack{m_1,\cdots, m_k\geq 0\\ m_1+2m_2+\cdots km_k=m'}}\ q^{h_{{\rm mwc(0)}}}  \prod _{j=1}^{k-1} \begin{bmatrix}
 p_j+m_j\\ m_j\end{bmatrix}\;.
 \end{equation} 
The label 0 reminds that  this expression refers to the  vacuum module. The finitized character of the charge-$r$ module is obtained by restricting the sum to terms with $ m'\equiv r $ mod $k$.

 The conformal character is reproduced in the  limit $m'\rw \y$ performed by taking $m_k\rw \y$. This implies that $p_j\rw \y$ for all $j$. Using
  \begin{equation}\lim_{m\rw\y} \begin{bmatrix}
m\\ n\end{bmatrix}= \frac{1}{(q)_n}\;,
\end{equation}  
this yields
\begin{equation} \label{ca0}\chi_0(q)= \sum_{m_1,\cdots, m_{k-1}\geq 0} 
\frac{ q^{h_{{\rm mwc(0)}}}  } {(q)_{m_1}\cdots (q)_{m_{k-1}}}\;. 
 \end{equation} 
 This is the expected result,  the Lepowski-Primc (vacuum) character \cite{LP}.
 Again, the character of the charge-$r$ submodule is obtained by restricting the sum to those terms which satisfy  $ m\equiv r $ mod $k$.

\section{Generalization to arbitrary irreducible modules}

An  irreducible module of the $\z_k$ parafermionic model is  characterized by its highest-weight state $|\varphi_\ell\R $, with $0\leq \ell \leq k-1$ and $|\varphi_0\R = |0 \R $. The charge of its highest-weight state  is $\q (|\varphi_\ell\R )=\ell$ and its conformal dimension, $h_\ell$, reads \cite{ZF}:
\begin{equation} h_\ell= \frac{\ell(k-\ell)}{2k(k+2)} \;. \end{equation}
Again each module can be separated into distinct submodules of fixed relative charge $2r$. The highest-weight state in the sector of relative charge $2r$ has conformal dimension \cite{JM1}:
\begin{equation} h_\ell^{(r)} = h_\ell + \frac{r(k-\ell-r)}{k} \;. \end{equation} 
The fractional dimension  of a state such as (\ref{stringext}) of charge $m'$  but with $|0\R $ replaced by  $| \varphi_\ell\R $ is
\begin{equation} h_{{\rm frac}} = \frac{m'(m'+\ell)}{k}\;.
\end{equation}
The corresponding RSOS path starts on the vertical axis at position $(0,\ell)$ and again it is forced to finish on the $x$ axis.  Denote by gsc($\ell, r)$ the ground-state configuration in the $r$-th sector  of the module specified by $\ell$. It is described as follows: the path starts with a peak of relative charge $r$ at position $r$, reaches the $x$-axis at position $2r+\ell$ and  it is  completed by a sequence of $m'-r$ peaks of charge $k$. This construction entails the following constraint:
\begin{equation} 0\leq \ell+r\leq k\;. 
\end{equation}
But recall that an independent set of highest-weight states of fixed charge  satisfies $0\leq r\leq k-\ell-1$ \cite{JM1}.  This is  more restrictive than the above condition, which therefore does not entail any loss of generality. It is simple to verify that 
\begin{equation} h= w_{{\rm gsc}(\ell, r)}- h_{{\rm frac}} +h_\ell=h_\ell^{(r)} \;. 
\end{equation}
 
 Consider now the modification of the minimal-weight configuration corresponding to a given set of values $m_j$, caused by changing  the value of the initial  vertex from $(0,0)$ to $(0,\ell)$. Let us denote this modified expression  by mwc($\ell$). This configuration is described as follows. The path starts at $y=\ell$ with a sequence of $m_1$ peaks of charge 1 (at positions $1,3,\ldots ,2m_1-1$ and note that the height of these peaks is $1+\ell$), followed by $m_2$ peaks of charge 2, etc. This proceeds until we reach the peaks of charge $k-\ell+1$. Since they would have height $k+1$, which is not allowed, the path, just after the sequence of $m_{k-\ell}$ peaks of charge $k-\ell$, must go down to  height $\ell-1$  before describing the closely-packed sequence of peaks of charge $k-\ell+1$. 
 The need for this extra SE edge increases the weight associated to these peaks of charge $k-\ell+1$ by $m_{k-\ell+1}$ since the peak  positions  are displaced by one unit each as compared 
 the minimal-weight configuration for $\ell=0$. Similarly, the peaks of charge $k-\ell+2$ must start at the height $\ell-2$ so that the straight-down segment following the  last peak of charge $k-\ell+1$, which has length $k-\ell+1$ in mwc(0), has now an extra SE edge. (See Fig. \ref{figure3} for an example with $\ell=2$ and $k=4$.) Comparing with  the minimal-weight configuration for $\ell=0$, there are two extra edges  before the sequence of peaks of charge $k-\ell+2$, so that all peak positions are displaced by two units; their contribution to the weight is thus augmented by $2 m_{k-\ell+2}$. A similar analysis applies to all higher charge peaks. The net result is that  
  \begin{equation} w_{{\rm mwc}(\ell)}= w_{{\rm mwc(0)}}+ m_{k-\ell+1}+2m_{k-\ell+2}+\cdots +\ell m_{k}.
\end{equation}

  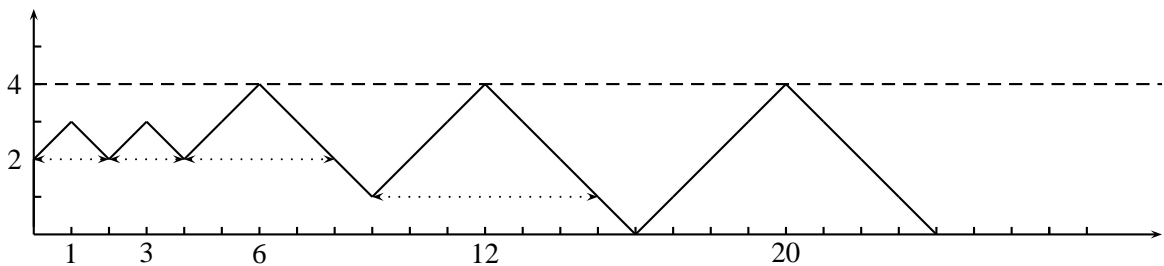
\begin{figure}[ht]
\caption{{\footnotesize The minimal-weight configuration for $k=4$ and $\ell=2$, with charge content $m_1=2$ and $ m_2=m_3=m_4=1$.}} \label{figure3}
\begin{center}
\begin{pspicture}(0,0)(15.5,4)
\psline{->}(0.5,0.5)(0.5,3.5) \psline{->}(0.5,0.5)(15.5,0.5)
\psset{linestyle=dashed,dashadjust=false}
\psline(0.5,2.5)(15.5,2.5)
\psset{linestyle=dotted}
\psline{<->}(0.5,1.5)(1.5,1.5)
\psline{<->}(1.5,1.5)(2.5,1.5)
\psline{<->}(2.5,1.5)(4.5,1.5)
\psline{<->}(5.0,1.0)(8.0,1.0)
\psset{linestyle=solid}
\psline{-}(0.5,1.5)(0.5,0.6) \psline{-}(1.0,0.5)(1.0,0.6)
\psline{-}(1.5,0.5)(1.5,0.6) \psline{-}(2.0,0.5)(2.0,0.6)
\psline{-}(2.5,0.5)(2.5,0.6) \psline{-}(3.0,0.5)(3.0,0.6)
\psline{-}(3.5,0.5)(3.5,0.6) \psline{-}(4.0,0.5)(4.0,0.6)
\psline{-}(4.5,0.5)(4.5,0.6) \psline{-}(5.0,0.5)(5.0,0.6)
\psline{-}(5.5,0.5)(5.5,0.6) \psline{-}(6.0,0.5)(6.0,0.6)
\psline{-}(6.5,0.5)(6.5,0.6) \psline{-}(7.0,0.5)(7.0,0.6)
\psline{-}(7.5,0.5)(7.5,0.6) \psline{-}(8.0,0.5)(8.0,0.6)
\psline{-}(8.5,0.5)(8.5,0.6) \psline{-}(9.0,0.5)(9.0,0.6)
\psline{-}(9.5,0.5)(9.5,0.6) \psline{-}(10.0,0.5)(10.0,0.6)
\psline{-}(10.5,0.5)(10.5,0.6) \psline{-}(11.0,0.5)(11.0,0.6)
\psline{-}(11.5,0.5)(11.5,0.6) \psline{-}(12.0,0.5)(12.0,0.6)
\psline{-}(12.5,0.5)(12.5,0.6) \psline{-}(13.0,0.5)(13.0,0.6)
\psline{-}(13.5,0.5)(13.5,0.6) \psline{-}(14.0,0.5)(14.0,0.6)
\psline{-}(14.0,0.5)(14.0,0.6)
\psline{-}(14.5,0.5)(14.5,0.6)\rput(1.0,0.25){{\small $1$}}
\rput(2.0,0.25){{\small $3$}} \rput(3.5,0.25){{\small $6$}}
\rput(6.5,0.25){{\small $12$}} \rput(10.5,0.25){{\small $20$}}
 \psline{-}(0.5,1.0)(0.6,1.0)
\psline{-}(0.5,1.5)(0.6,1.5) \psline{-}(0.5,2.0)(0.6,2.0)
\psline{-}(0.5,2.5)(0.6,2.5) \psline{-}(0.5,3.0)(0.6,3.0)
\rput(0.25,1.5){{\small $2$}} \rput(0.25,2.5){{\small $4$}}
\psline{-}(0.5,1.5)(1.0,2.0) \psline{-}(1.0,2.0)(1.5,1.5)
\psline{-}(1.5,1.5)(2.0,2.0) \psline{-}(2.0,2.0)(2.5,1.5)
\psline{-}(2.5,1.5)(3.0,2.0) \psline{-}(3.0,2.0)(3.5,2.5)
\psline{-}(3.5,2.5)(4.0,2.0) \psline{-}(4.0,2.0)(4.5,1.5)
\psline{-}(4.5,1.5)(5.0,1.0) \psline{-}(5.0,1.0)(5.5,1.5)
\psline{-}(5.5,1.5)(6.0,2.0) \psline{-}(6.0,2.0)(6.5,2.5)
\psline{-}(6.5,2.5)(7.0,2.0) \psline{-}(7.0,2.0)(7.5,1.5)
\psline{-}(7.5,1.5)(8.0,1.0) \psline{-}(8.0,1.0)(8.5,0.5)
\psline{-}(8.5,0.5)(9.0,1.0) \psline{-}(9.0,1.0)(9.5,1.5)
\psline{-}(9.5,1.5)(10.0,2.0)
 \psline{-}(10.0,2.0)(10.5,2.5)\psline{-}(10.5,2.5)(11.0,2.0)
\psline{-}(11.0,2.0)(11.5,1.5) \psline{-}(11.5,1.5)(12.0,1.0)
\psline{-}(12.0,1.0)(12.5,0.5) 

\end{pspicture}
\end{center}
\end{figure}

 The sole effect of this change on the character is to replace $h_{{\rm mwc(0)}}$ in (\ref{ca0}) by $h_{{\rm mwc}(\ell)}$ with
 \begin{equation} h_{{\rm mwc}(\ell)}= w_{{\rm mwc}(\ell)}- \frac{m'(m'+\ell)}{k}\;.
\end{equation}
The finitized character of the module with highest-weight state $ |\varphi_\ell\R$ is therefore
 \begin{equation}\label{rsosca}
 \chi^{(m')}_\ell(q) = \sum_{\substack{m_1,\cdots, m_k\geq 0\\ m_1+2m_2+\cdots km_k=m'}}\ q^{h_{{\rm mwc}(\ell)}}  \prod _{j=1}^{k-1} \begin{bmatrix}
 p_j+m_j\\ m_j\end{bmatrix}\;.
 \end{equation} 

Finally, we indicate the  modifications on the conditions defining $\Gamma^{[k]}$ (trivially adapted from those pertaining to $\La^{[k-1]}$ given in \cite{JM.A}): the condition (\ref{diCg}) is unchanged but (\ref{loBg}) and (\ref{upg}) are modified as follows
\begin{equation}\label{loBgl}
   \g^{(j)}_{m_j} \geq j+{\rm max}\, (j+\ell-k,0)+
2j (m_{j+1}+\cdots +  m_{k}), 
\end{equation} 
 and 
 \begin{equation} \label{upgl}
\g^{(j)}_1 \leq -j  +{\rm max}\, (j+\ell-k,0)+ 2\,[\, jm_j+(j+1)m_{j+1}+\cdots +km_k\,]\;,
\end{equation} 
 respectively. The vertical  position of the initial  vertex becomes max $(\ell-1,0)$. 
 
\section{Concluding remarks}

The states in irreducible parafermionic modules are known to be  described by two different types of paths. The main result of this work is the presentation of a weight-preserving bijection between these two path descriptions. Both representations lead to the same fermionic expression for the conformal characters. Indeed, in the RSOS case, this is the expression (\ref{rsosca}) evaluated in the limit $m_k\rw\y$ (and again, $m_k$ cancels in $h_{{\rm mwc}(\ell)}$). On the other hand, the generating function for the Bressoud paths, with the weight adjusted to take into account the fractional part of the parafermionic modes,  is precisely this limiting expression \cite{BreL, JM.A}. Therefore, even if we have two different descriptions of the space of states, we end up with a single fermionic  expression for the conformal characters. However, it should be stressed that the RSOS formulation  provides a finitization of these fermionic forms, which is not the case for the other description since the Bresssoud paths do not embody a natural notion of length. A short self-contained derivation of these finite versions has been presented in Sections 3 and 4.


 \vskip0.3cm
\noindent {\bf ACKNOWLEDGMENTS}

PJ would like to  thank P. Pearce and J. Rasmussen for fruitful  discussions as well as the organizers of the workshop {\it From statistical mechanics to conformal and quantum field theory} for their hospitality. We also thank D. Ridout for a very critical reading of the manuscript and useful suggestions.
 The work of PJ is supported by EPSRC, while that of  PM is supported  by NSERC.


\end{document}